\documentclass[amsmath,amssymb,aps,reprint,nofootinbib]{revtex4-2}

\usepackage{physics}
\usepackage{tensor}

\begin{document}

\title{A Purely Relativistic Point-Source Boundary Condition for the Schwarzschild Solution}

\author{Peter Hayman}
\email[]{peter@haymanphysics.com}
\affiliation{Department of Physics, The University of Auckland}

\date{\today}

\begin{abstract}
    We present a simple derivation of a point-source boundary condition for the Schwarzschild solution that relates the Schwarzschild radius to the mass of its source without appealing to the Newtonian limit. Interpretation of the Schwarzschild radius in terms of the mass of a point-like source traditionally means resorting to distant asymptotics and the safety of Newtonian gravity, but here we instead show a direct connection between a point-particle's invariant mass and the length parameter of the Schwarzschild solution it sources, fully within the framework of general relativity. As a corollary, we also explain why attempts to show this by distributional techniques often result in a physically unmotivated spatial distribution for the source stress-energy tensor.
\end{abstract}

\maketitle

\section{Introduction}
Any introductory textbook on general relativity provides a similar derivation of the Schwarzschild black hole solution, the logic of which is as follows. It is said that we aim to study the spacetime geometry outside a finite-sized spherically symmetric source, so the most general static spherically symmetric metric is solved for through the vacuum Einstein field equations, and two physical boundary conditions are imposed. First, the solution is required to be asymptotically Minkowski which imposes $g_{tt}g_{rr} = 1$ globally (though we will show here that this actually fails to hold exactly at the source), and second that it slightly less asymptotically reduces to the Newtonian result for a gravitational field outside a finite-sized mass distribution\footnote{Defining the same parameter by using a model of compact stress-energy density a la \cite{schutz_2009} ultimately faces the same challenge, an integral over the model $T\indices{_0^0}$ is named ``$M$'' and matched onto the Newtonian solution for good measure.}. This second condition is problematic in that it interprets the mass parameter of the Schwarzschild solution (i.e., the $M$ in $r_s = 2GM$) by \emph{analogy} with a purely Newtonian system, rather than directly connecting it to the source of stress-energy that purportedly generated the solution in the first place. Even formalizing the analogy by using the ADM technique still evades the fundamental problem of connecting the intrinsic length scale of the Schwarzschild geometry to the intrinsic energy scale of the stress-energy tensor that must be its source.  

Of course this conceptual shortcut does not hamper practical astrophysical research that generally keeps close enough to the Newtonian regime anyway, but it should be a thorn in the side of any theoretician who enjoys the self-consistency of general relativity, even if only as an effective theory. And indeed the literature has not been silent on the issue, even going back as early as Einstein and Rosen in 1935 \cite{einstein_particle_1935}. More recent work (such as \cite{balasin_energy-momentum_1993,kawai_distributional_1997,pantoja_energy_1997,heinzle_remarks_2002,pantoja_distributional_2002,petrov_schwarzschild_2005,steinbauer_use_2006,petrov_point_2018}) has made very interesting advances using modern distribution theory, but going the wrong logical direction---starting with the Schwarzschild solution and arguing that it leads to a point-source stress-energy via the Einstein equations. Taking this route it can certainly be shown that the source of the Schwarzschild solution must be a point-energy density, but in principle these actually just show the energy density has scale $|T| \sim r_s/2G$, not the other way around. Moreover, a number of these calculations even return a stress-energy tensor with physically unmotivated spatial components in the usual Schwarzschild coordinates. Some works of course do go the correct logical direction \cite{fiziev_solutions_2004, katanaev_point_2013}, but they do so in adventurous ways, exploring wider solution spaces with advanced techniques.


Here we show that a middle ground really does exist, one can set the problem up with a sensible relativistic point-source of stress energy and use it to generate a near-source boundary condition that fixes the Schwarzschild radius parameter in terms of the invariant rest mass of the source, all at a level of rigour that preserves the simplicity of the Schwarzschild solution. 

\section{Setup}

We set the problem up as follows. The aim is to solve the Einstein field equations now coupled to the geodesic equation:
\begin{equation}
    \label{eq:EOMs}
    G_{\mu\nu} = \kappa T^\text{pp}_{\mu\nu}, \qquad \text{and} \qquad U^\nu \nabla_\nu U^\mu = 0,
\end{equation}
where the Einstein tensor is $G_{\mu\nu} = R_{\mu\nu} - \frac 12 g_{\mu\nu} R$ . The stress-energy is best found from a variational principle, but it is easy enough to argue on dimensional and tensorial grounds that it must have the putative form:
\begin{equation}
    \label{eq:SETput}
    T^\text{pp (put.)}_{\mu\nu} := \int \dd\tau \frac{M}{\sqrt{-g}} U_{\mu}(\tau) U_{\nu}(\tau) \delta^{(4)}(x^\mu - x^\mu_0(\tau)),
\end{equation}
where $\tau$ is the proper time parameterizing the source worldline, $M$ is its invariant mass $M := \sqrt{-p_\mu p^\mu}$, $x^\mu_0(\tau)$ is the source worldline and $U^\mu(\tau) = \dv[]{x^\mu}{\tau}$ its four-velocity. 

Here is where the first subtlety arises; whatever the form of the worldline solution $x^\mu_0(\tau)$, it is a one-dimensional path in spacetime, so we can always define one coordinate $t$ to run along the path, and the others such that the otherwise empty spatial geometry is spherically symmetric about $x^i(\tau)$. Of course we know this solution, it is just the Schwarzschild geometry, but with that comes the metric singularity at $x^i(\tau)$, and importantly that means $U^\mu(\tau) = g^{\mu\nu}(x^\mu(\tau))U_\nu(\tau)$ is undefined. Since in this case the inverse metric is evaluated exactly on the singular point, it is impossible to resolve this by an approximation or limiting procedure, so the solution must be to separately define $U_\mu := (-1,\,0,\,0,\,0)$ and $U^\mu = (1,\,0,\,0,\,0)^T$, and pick one type of stress-energy tensor to work with and stick with that choice. This amounts to inputting the source four-velocity and its dual as an external parameter, but that is nothing new in general relativistic problems. This is a known issue and the conventional answer (which we follow) is to use the mixed form \cite{balasin_energy-momentum_1993} since it also allows an unambiguous calculation of the trace of the stress-energy if necessary. Thus the problem is better phrased as seeking the solution to the field equations in mixed form
\begin{equation}
    \label{eq:mixedEFE}
    R\indices{_\mu^\nu} - \frac 12 \delta_\mu^\nu R = \kappa T\indices{_\mu^\nu},
\end{equation}
with stress-energy tensor
\begin{equation}
    \label{eq:SET}
    T\indices{_\mu^\nu} := -\frac{M}{\sqrt{-g}}  \delta_\mu^0 \delta^\nu_0 \delta^{(3)}(\vec{x}).
\end{equation}

Everywhere away from the source, the stress-energy vanishes and the metric can be put in the Schwarzschild form:
\begin{equation}
    \label{eq:genSch}
    \dd s^2 = -q(r) \dd t^2 + f(r) \dd r^2 + r^2 \dd\Omega^2,
\end{equation}
where the functions $q(r)$ and $f(r)$ are allowed to take any sign as we are solving the geometry of the entire spacetime away from the source singularity. The next subtlety we run into is the choice of coordinates. The usual Schwarzschild coordinates $(t,r,\theta,\phi)$ are very convenient away from the source, but they are not actually defined at the source itself, so in order to include the stress-energy \eqref{eq:SET}, we need to use a coordinate system that covers the origin. We find the simplest choice is a pseudo-Cartesian system defined by the transformation (see e.g.~\cite{kawai_distributional_1997})
\begin{equation}
    \label{eq:defCart}
    \begin{aligned}
        \hat{x} := r\sin\theta\cos\phi, &\qquad \hat{y} := r\sin\theta\sin\phi,\\
        \qquad \text{and} \qquad \hat{z} &:= r\cos\theta.
    \end{aligned}
\end{equation}
(Note that these are only superficially Cartesian coordinates, they are not properly so since they do not describe a Euclidean space). In these coordinates, the Einstein field equations nicely work out to (using hats for the components in Cartesian coordinates)
    \begin{align}
        \hat{R}\indices{_t^t} &= R\indices{_t^t}, \quad \text{and}  \label{eq:cartRicT} \\
        \hat{R}\indices{_j^i} &= \frac{\hat{x}^i \hat{x}^j}{r^2}R\indices{_r^r} + \left( \delta^i_j - \frac{\hat{x}^i\hat{x}^j}{r^2} \right)R\indices{_\theta^\theta} \label{eq:cartRicSp}
    \end{align}
    in terms of the (hatless) Schwarzschild coordinate terms
    \begin{align}
        \label{eq:polMix}
        R\indices{_t^t} &= -\frac {q^{\prime\prime}}{2fq} + \frac 1 4 \frac {q^\prime}{fq}\frac {(fq)^\prime}{fq} - \frac 1 {rf} \left( \frac {q^\prime} q \right),  \\
        R\indices{_r^r} &= -\frac {q^{\prime\prime}}{2fq} + \frac 1 4 \frac {q^\prime}{fq} \frac{(fq)^\prime}{fq} + \frac 1 {rf} \left( \frac {f^\prime} f \right), \quad \text{and} \\
        R\indices{_\theta^\theta} &= \frac 1 {r^2} \partial_r\left( r\left(1 - \frac 1 f\right) \right) - \frac 12 \frac 1 {rf} \frac{(fq)^\prime}{fq},
    \end{align}
    and where primes denote differentiation with respect to $r$.

    Even before computing the Einstein tensor, the absence of off-diagonal components of $T\indices{_\mu^\nu}$ combined with \eqref{eq:cartRicSp} implies $R\indices{_r^r} = R\indices{_\theta^\theta}$ exactly. Moreover, by inspection $R\indices{_t^t} = R\indices{_r^r} - \frac 1 {rf}\frac{(fq)^\prime}{fq}$, so the Einstein equations cleanly reduce to
    \begin{align}
        -R\indices{_\theta^\theta} - \frac 12 \frac 1 {rf}\frac{(fq)^\prime}{fq} &= -\frac{M\kappa}{\sqrt{-g}} \delta^{(3)}(\vec{x}), \quad \text{and} \label{eq:cartG1}
\\
        -R\indices{_\theta^\theta} + \frac 12 \frac 1 {rf} \frac{(fq)^\prime}{fq} &= 0. \label{eq:cartG2}
    \end{align}
    Away from the source, the solutions are trivially the usual Schwarzschild solutions, 
    \begin{equation}
        \label{eq:extSols}
        f(r)q(r) = 1, \qquad \text{and} \qquad f(r) = \frac 1 {1 - r_s/r},
    \end{equation}
    using the asymptotic Minkowski boundary condition to fix the constant in the first equation, and leaving the integration constant in the second as yet undetermined (though suggestively named).

\section{Near-Source Boundary Condition}

    Including the source, the Einstein equations can be arranged to find
    \begin{equation}
        \label{eq:prepBC}
        \frac 1 {r^2} \partial_r \left( r\left( 1 - \frac 1 {f(r)} \right)  \right) = \frac {M\kappa}{\sqrt{-g}} \delta^{(3)}(\vec{x}),
    \end{equation}
    however the solution $f(r)$ must be handled with care here. It is well known (see e.g.~\cite{balasin_energy-momentum_1993, kawai_distributional_1997}) that the Schwarzschild metric components are too singular and need to be regularized near the origin to handle their delta function source, so before integrating \eqref{eq:prepBC}, introduce a regulating function $h(r;\epsilon)$ to  $f(r)$:
    \begin{equation}
        \label{eq:regF}
        f_\epsilon := \frac 1 {1 - h(r;\epsilon)r_s/r}.
    \end{equation}
    The regulator $h(r;\epsilon)$ can be very general and only needs to satisfy that $\lim_{\epsilon \to 0} h(r;\epsilon) \to 1$ , and that $h(0;\epsilon) = 0$ at least as fast as $r^2$. One example would be $h(r;\epsilon) = r^2/(r^2 + \epsilon^2)$, but any similar function will work just as well. The near-source boundary condition is then found by integrating \eqref{eq:prepBC} over a Euclidean volume\footnote{Here we are thinking of \eqref{eq:prepBC} as a simple equality between functions (or distributions) on $\mathbb{R}^3$. If you find this unsatisfactory, just integrate over a small enough volume that the regulated $\sqrt{g_3} = \sqrt{f(r)} \sim \text{const}$. If the role-reversal of $t$ and $r$ inside the Schwarzschild horizon is a concern, perform the whole calculation for a particle of mass $-M$ instead and extrapolate the result to a real source.} and then taking the limit $\epsilon\to 0$. 
    \begin{equation}
        \label{eq:protoBC}
        4\pi r_s \lim_{\epsilon \to 0} h(r;\epsilon)\big\vert^R_0 = 4\pi r_s = M\kappa,
    \end{equation}
    from which we find the correct identification $r_s = 2GM$ (using $\kappa := 8\pi G$).

    Notice that consistency of the Einstein equations \eqref{eq:cartG1} and \eqref{eq:cartG2} also requires
    \begin{equation}
        \label{eq:prepBC2}
        \frac 1 {rf} \frac{(fq)^\prime}{fq} = \frac {M\kappa}{\sqrt{-g}} \delta^{(3)}(\vec{x}).
    \end{equation}
    Coupled with \eqref{eq:regF} and \eqref{eq:protoBC} this boundary condition fixes $q(r)$ at the origin, and while the explicit solution is not important, what matters is that it is distinctly \emph{not} $f(r)q(r) = 1$. Artificially imposing that $f(r)q(r) = 1$ exactly at the origin implies the source stress-energy must have had (physically unjustified) additional components that cancel out the right-hand side of \eqref{eq:prepBC2}.

\section{Discussion}

We have presented a simple yet subtle derivation of the near-source boundary condition that identifies the Schwarzschild radius of a black hole with the invariant rest mass of the point-source that generates its geometry, all entirely within the framework of general relativity. This boundary condition is physically intuitive and does not invoke or propose any more complications than strictly necessary. Importantly, it provides confirmation that from first principles the mass found in the Schwarzschild solution really does coincide with the rest mass of a point-particle source, and that this statement holds purely within the framework of general relativity without appeal to an asymptotic Newtonian limit. The regularization of the metric that was necessary at the source does not prevent but rather supports this conclusion; it is standard practice to interpret infinite-density point sources as effective descriptions of more complicated physics, as for example the Coulomb potential in electrostatics is only an effective description of finite-sized sources. In this way, our solution serves as a general relativistic analog to Gauss's law in electrostatics, in this case saying that the geometry outside any finite, static source of stress-energy is equivalent to that generated by a pure point-source.  

An interesting corollary is that this approach also settles a curious issue that is often seen in the literature on this topic. In a number of cases (for example \cite{balasin_energy-momentum_1993, pantoja_energy_1997}), it is found that the source of stress energy in Schwarzschild coordinates has to have a rather peculiar and physically unmotivated form, $T\indices{_\mu^\nu} = (-\frac{M}{\sqrt{-g}}  \delta^{(3)}(\vec{x}))\text{diag}(1,\,1,\,-\frac 12,\,-\frac 12)$. As noted above, this is an artifact of their logic of starting from the Schwarzschild solution with $g_{tt}g_{rr} = 1$ exactly as opposed to using the point-source to derive the Schwarzschild solution. Starting instead from the physically-motivated stress energy as we do, the correct statement is that $g_{tt}$ and $g_{rr}$ deviate from each other at the source through the boundary condition \eqref{eq:prepBC2}. We will also note that while most works simply state this as a curiosity or otherwise bypass it, the authors of \cite{pantoja_distributional_2002} incorrectly state that the two stress-energies are exactly equivalent.

Finally, we address a common alternative viewpoint that all of this is unnecessary, that Schwarzschild black holes are pure vacuua and the world just happens to be missing a point. To some extent, the calculation above is sufficient to dismiss this on the grounds that if a stress-energy can be found to generate this geometry, then it \emph{must} generate this geometry by virtue of the Einstein field equations. We will just add a few more arguments to support this. First, permanent removal of a point in the Schwarzschild solution is inconsistent with its interpretation as the causal future geometry of an extended stress-energy distribution (such as a star) in which that point was considered a part of spacetime. Second, removing the singularity from general relativity and assigning it instead to some as-yet unknown superceding theory still doesn't make the Schwarzschild geometry a pure vacuum, it just passes the buck to a new theory to define the physical meaning of the scale $r_s$ by means of a correspondence principle (which in this case would be the correct logical direction, but is impossible to do without knowledge of that new theory). Furthermore, the previous calculations of the distributional limit of the Einstein field equations support our view as well; if the geometry side of the field equations for the Schwarzschild metric necessarily approach the form of the stress-energy of a point-source, it is a challenge indeed to insist it is all still just a grand coincidence.


\section{Acknowledgements}

Thanks to Richard Easther and Jens Niemeyer for helpful discussions. This work is supported by the Marsden Fund of the Royal Society of New Zealand. I am also very grateful to Richard Easther and the Physics department at The University of Auckland for giving me the opportunity to teach general relativity and learn a surprising number of interesting subtleties that lurk in a logically consistent, cohesive narrative introduction to the topic.

\bibliography{gr}

\end{document}